# Design of a High-Power and High-Efficiency GaN-HEMT VCO Based on an Inverse Class-F Amplifier

Junlin Mi, Ruinan Fan, *Student Member, IEEE*, Liping Yan, *Senior Member, IEEE*, Yuhao Feng, and Changjun Liu, *Senior Member, IEEE*

*Abstract*— This letter proposes a high-power and high-efficiency GaN-HEMT voltage-controlled oscillator (VCO). The VCO consists of a coupled-line coupler, an inverse class-F amplifier, and a novel frequency-tunable stepped-impedance resonator (SIR). Using a harmonic control circuit and a parasitic parameter compensation circuit, the power amplifier (PA) operates in the inverse class-F state to achieve high efficiency. The feedback circuit uses a coupled-line coupler instead of the traditional coupling capacitor to control feedback power precisely. The measurement results show that the VCO with an oscillation frequency of 2.41–2.45 GHz achieves a maximum conversion efficiency of 74.5% at 2.44 GHz and an output power of 40.2 dBm. It is a candidate for the microwave source in a wireless power transmission system.

## I. Introduction

AS an indispensable part of the microwave wireless power transmission (MWPT) system, microwave power sources significantly impact the overall system performance. Currently, the commonly used microwave power sources are magnetrons, conventional amplifier-type microwave generators, and solid-state power oscillators. Conventional amplifier-type microwave generators require multistage amplification, which leads to a larger circuit size, reduced efficiency, and increased cost [1]. GaN HEMTs are widely used in designing high-efficiency RF power amplifiers (PAs) and power oscillators due to their high breakdown voltage and power density at microwave frequencies. Compared to magnetrons, the solid-state power oscillators designed based on GaN power devices have advantages, such as longer lifespan, higher reliability, and lower power supply voltage [2].

Received 13 October 2024; accepted 5 December 2024. This work was supported in part by NSFC under Grant U22A2015 and Grant 62071316 and in part by Sichuan Science and Technology Program under Grant 2024YFHZ0282. *(Corresponding author: Changjun Liu.)*

Junlin Mi, Ruinan Fan, Liping Yan, and Changjun Liu are with the School of Electronics and Information Engineering, Sichuan University, Chengdu 610064, China, and also with Yibin Industrial Technology Research Institute of Sichuan University, Yibin 644000, China (e-mail: cjliu@ieee.org).

Yuhao Feng is with the School of Electronic Science and Engineering, University of Electronic Science and Technology of China, Chengdu 611731, China.

Color versions of one or more figures in this letter are available at https://doi.org/10.1109/LMWT.2024.3514739.

Digital Object Identifier 10.1109/LMWT.2024.3514739

The efficient harmonic-tuned GaN power oscillators have been reported in [3], [4], and [5]. However, they lack frequency tunability. Injection locking can tune frequency by a small reference signal, but the tuning range is very narrow, such as injection-locked magnetrons [6], [7], [8], [9], [10]. Frequency-tunable high-power GaN oscillators have yet to be widely studied so far. In [11] and [12], the frequency-tunable high-power GaN oscillators with mechanical phase shifters in the feedback have been reported. In [13], with the combination of a ceramic coaxial resonator, coupling capacitor, and a multisection varactor diode, a GaN-HEMT voltage-controlled oscillator (VCO) having an output power of 20 W and a drain efficiency of 55% has been successfully achieved. However, the feedback amplitude and phase from coupling capacitors cannot be precisely controlled. In addition, the ceramic coaxial resonator and coupling capacitors must be used, resulting in more parasitic parameters in the feedback circuit.

In a WPT system, efficiency is a crucial metric [14], [15]. The class F/F$^{-1}$ PA has received significant attention due to its simple structure, easy implementation, and theoretical maximum efficiency of 100% [16], [17], [18], [19], [20], [21], [22], [23]. An inverse class-F amplifier is realized in the VCO circuit to guarantee high efficiency. A coupled-line coupler has been added instead of a traditional coupling capacitor to control the amplitude and phase of the feedback power precisely. Furthermore, a novel frequency-tunable stepped-impedance resonator (SIR) is used in the feedback path to change the oscillation frequency. Experimental results agree with the simulations from the harmonic balance simulator obtained in the advanced design system (ADS).

This letter reports on a high-efficiency 2.45-GHz GaN VCO fabricated on F4B with a tuning range of 40 MHz, a maximum conversion efficiency of 74.5%, and an output power of 40.2 dBm at a supply $V_{ds} = 30$ V and $V_{gs} = -2.6$ V.

## II. Design of an Inverse Class-F Amplifier

Since efficiency is a crucial metric in WPT systems, a high-efficiency inverse class-F PA with a GaN HEMT CGH40010F is employed. The GaN HEMT has extrinsic parasitic components caused by the package, interconnection, and bonding wires. The parasitic elements provided by the manufacturer are $C_{ds} = 1.2$ pF, $C_p = 0.2$ pF, and $L_d = 0.55$ nH. To satisfy the harmonics impedance conditions of an inverse class-F PA at its intrinsic drain [$Z_{in}(2\omega_0) = \infty$, $Z_{in}(3\omega_0) = 0$, $Z_{in}$ is the input impedance, and $\omega_0$ is a fundamental frequency], the output matching circuit shown in Fig. 1 consists of a harmonic tuning network, a compensation network, and a fundamental frequency matching network.

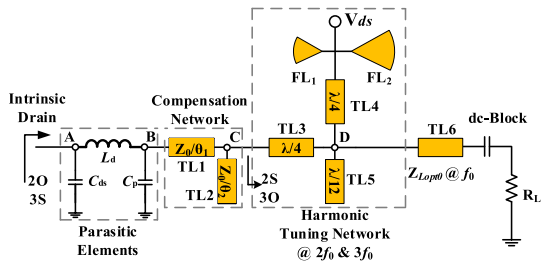

Fig. 1. Output matching circuit with parasitic compensation for the inverse class-F amplifier ($\lambda$ is the fundamental wavelength, $f_0$ is the fundamental frequency of 2.45 GHz, $Z_0$ is the intrinsic impedance, and $\theta$ is the electrical length).

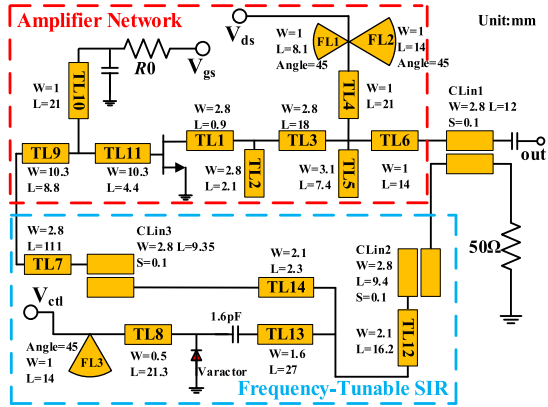

Fig. 2. Schematic of the proposed VCO with an amplifier, a frequency-tunable SIR, and a coupled-line coupler.

In the harmonic tuning network, the $\lambda/12$ open stub TL$_5$ creates a short circuit at node D at $3f_0$ (7.35 GHz). Through a $\lambda/4$ transmission line TL$_4$, fan-shaped FL$_1$ and FL$_2$ transform the impedance at node D to the open at $f_0$ (2.45 GHz) and the short at $2f_0$ (4.9 GHz). As a result, the subsequent fundamental matching network TL$_6$ cannot affect the harmonics impedance at point D at $3f_0$ and $2f_0$. TL$_3$, with the length of $\lambda/4$, is three times a quarter wavelength for the third harmonics, which presents with infinite impedance at $3f_0$ at node C.

Furthermore, the electrical length of TL$_3$ is half a wavelength for the second harmonic, which presents zero impedance at $2f_0$ at node C. An L-shaped parasitic compensation circuit [24] transforms impedance to satisfy the class-F$^{-1}$ harmonics impedance conditions at the intrinsic drain.

The load–pull and source–pull simulations are carried out at 2.45 GHz in ADS with the harmonic tuning and compensation networks. At a drain bias voltage of 28 V and a gate bias voltage of $-2.8$ V, the subsequent output fundamental impedance $Z_{Lopt0}$ and input fundamental impedance are $105.6 + j52$ $\Omega$ and $8.8 - j6.4$ $\Omega$, respectively.

The whole structure of the amplifier network as part of the high-power VCO is shown in Fig. 2. Based on the harmonic balance simulator, the output power is 40.9 dBm, the gain is 12.9 dBm, and the power-added efficiency (PAE) is 82.6%, with the input power of 28 dBm at 2.45 GHz.

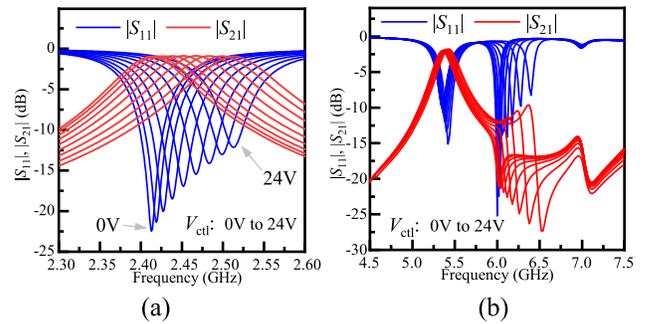

Fig. 3. Tunable SIR simulation results. (a) Simulation in its operating bandwidth. (b) Simulation at $2f_0$ and $3f_0$.

## III. FEEDBACK CIRCUIT WITH A TUNABLE SIR

It is necessary to design a frequency-tunable resonator to make the oscillator's frequency tunable. SIR boasts benefits, including a compact design, high $Q$-factors, and superior harmonic suppression [25]. Utilizing a frequency-tunable stub-loaded SIR in the feedback loop, as depicted in Fig. 2, effectively mitigates harmonics and is integral to the VCO's functionality. The resonator is coupled to the circuit by parallel lines. A varactor BB555 from Infineon is employed in the design. Its series capacitance tuning range is from 2 to 19 pF when the bias voltage changes from 30 to 0 V. A stub is loaded at a $\lambda/2$ SIR, and a varactor BB555 has been added at the end of the loaded stub. A 1.6-pF Murata capacitor is in front of the varactor as dc blocking and to change the capacitance collaboratively.

The proposed frequency-tunable SIR simulation results in its operating bandwidth are shown in Fig. 3(a). The varactor's capacitance is tuned by changing its reverse bias voltage $V_{ctl}$ to adjust the resonant frequency. While the bias voltage $V_{ctl}$ varies from 0 to 24 V, the resonant frequency tunes from 2.41 to 2.51 GHz, $|S_{21}|$ is about $-0.9$ dB, and $|S_{11}|$ is less than $-10$ dB in the band. Fig. 3(a) reveals that the resonator exhibits a $Q$-factor of approximately 26 across the bandwidth. Fig. 3(b) demonstrates the resonator's harmonic suppression capabilities at $2f_0$ and $3f_0$.

The power oscillator composed of the amplifier and the frequency-tunable network ought to satisfy (1) and (2) from Barkhausen criteria [26]

$$A(s) \cdot \beta(s) \geq 1 \tag{1}$$

$$\sum \Phi = 2\pi \times n, \quad (n = 1, 2, \ldots). \tag{2}$$

Part of the output power of the PA enters the transistor input through the feedback loop. The feedback power should be low enough not to degrade the output power and efficiency as long as the negative gain $\beta(s)$ of feedback satisfies (1). A coupled-line coupler replaces the traditional coupling capacitor to adjust the feedback power more accurately, as shown in Fig. 4(a). Port 1 is the PA output and the input of the coupler, port 2 is the transmitted port as the oscillator output, port 3 is the coupling port, and port 4 is the isolation port. Furthermore, compared to coupling capacitors, the experimental error of using coupled-line couplers is less, and fewer lumped components need to be soldered.

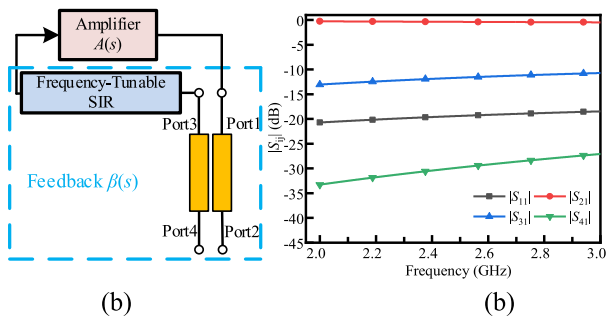

Fig. 4. (a) Coupled-line coupler in the feedback loop. (b) Simulation results of the coupled-line coupler.

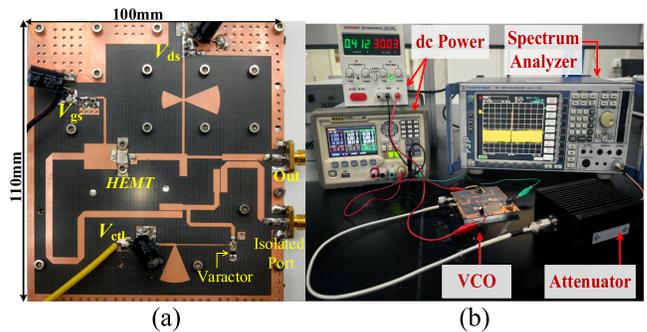

Fig. 5. (a) Fabricated prototype VCO. (b) Measurement setup for the proposed VCO.

The proposed class-F$^{-1}$ PA gets maximum efficiency at an output power of 40.9 dBm, while the input power is 28 dBm. Therefore, $\beta(s)$ of the feedback circuit should be −12.9 dB to ensure that the VCO gets the same performance as the PA. $|S_{21}|$ of the proposed frequency-tunable SIR is about −0.9 dB at the resonant frequency. Consequently, the feedback quantity $|S_{31}|$ of the coupled-line coupler should be about −12 dB. In addition, $|S_{21}|$ of the coupler should be close to zero to reduce the influence of the coupler on the output power.

The $S$-parameter simulation results of the coupled-line coupler are shown in Fig. 4(b). $|S_{21}|$ is close to zero, and $|S_{31}|$ is almost −12 dB around 2.45 GHz, which satisfies the design requirements of the feedback network. The optimized schematic of the proposed high-power VCO with an amplifier network, a coupled-line coupler, and a frequency-tunable SIR has been shown in Fig. 2, where CLin$_1$ is the designed coupled-line coupler, and its output impedance is 50 Ω to match the load. To meet the phase condition from Barkhausen criteria, TL$_7$ is a 50-Ω transmission line to compensate for the phase in the feedback loop.

## IV. IMPLEMENTATION AND EXPERIMENTAL RESULTS

The fabricated prototype VCO is shown in Fig. 5(a). The proposed VCO was mounted on the F4B board (110 × 100 mm) at a thickness of 1 mm. Its relative dielectric constant and the loss tangent are 2.65 and 0.002, respectively. An 8.2-pF Murata capacitor was used as dc blocking. The measurement setup is shown in Fig. 5(b). The bias voltages of the drain and source were supplied by dc power MS-303DS and DP1308A, respectively. An R&S FSP40 spectrum analyzer measured the output signal. A 30-dB attenuator protected the spectrum analyzer. The comparison between the simulation and measurement of the VCO as a function of frequency has been shown in Fig. 6. The simulated VCO had an output power of 39.2–40.2 dBm and a drain efficiency of 62.5%–75.5% between 2416 and 2470 MHz. The fabricated VCO had an output power of 38–40.2 dBm and a drain efficiency of 53%–74.5% with a drain bias voltage of 30 V and a gate bias voltage of −2.6 V between 2412 and 2453 MHz. The maximum drain efficiency was 74.5% and an output power of 40.2 dBm. Compared to the simulation results, the measured bandwidth of the VCO is comparatively narrow. This may be due to the suboptimal tuning range of the actual varactor.

Table I shows the comparison between the proposed VCO and previous work. In an earlier work, ceramic coaxial resonators and coupling capacitors must be soldered, resulting in more parasitic parameters in the feedback circuit. In addition, it is difficult to control the amplitude and phase of feedback power using coupling capacitors and mechanical phase shifters, resulting in a more complex system. This work offers several advantages over prior work, including fewer lumped components, more precise feedback control, enhanced efficiency, and a design entirely based on microstrip lines.

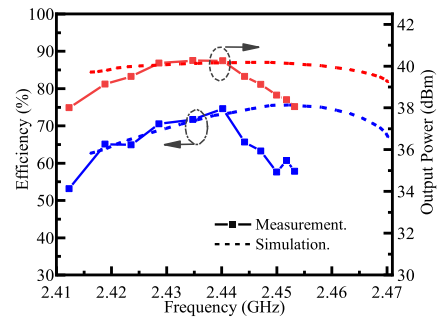

Fig. 6. Measured and simulated results of the proposed VCO.

TABLE I
PERFORMANCE WITH THE PRIOR HIGH-POWER VCO

| Ref. | Frequency selective network | Feedback | $f$ (MHz) | $P_{out}$ (dBm) | $Eff.$ (%) | Range (MHz) |
|---|---|---|---|---|---|---|
| [6] | mechanical phase shifter | capacitor | 920 | 43.5 | 67 | 60 |
| [12] | mechanical phase shifter | capacitor | 915 | 52 | 56 | 26 |
| [13] | Ceramic resonator with varactors | capacitor | 2450 | 43 | 55 | 71 |
| [27] | LC with varactors | capacitor | 2450 | 47 | 45 | 100 |
| This work | microstrip SIR with varactors | microstrip coupler | 2450 | 40.2 | 75 | 40 |

## V. CONCLUSION

In this letter, a high-efficiency and high-power VCO is investigated. The VCO consists of a coupled-line coupler, an inverse class-F amplifier, and a frequency-tunable SIR. Compared to previous work, this work has advantages, such as fewer lumped components, more accurate feedback control, higher efficiency, and is entirely composed of microstrip lines. The VCO has a tuning range of 40 MHz, a maximum conversion efficiency of 74.5%, and a maximum output power of 40.2 dBm. The proposed design method can be applied to solid-state microwave power sources for WPT applications.

## References

[1] H. Ikeda and Y. Itoh, "2.4-GHz-band high-power and high-efficiency solid-state injection-locked oscillator," *IEEE Trans. Microw. Theory Techn.*, vol. 66, no. 7, pp. 3315–3322, Jul. 2018.

[2] H. Ikeda and Y. Itoh, "A 2.4 GHz-band 250 W, 60% feedback-type GaN-HFET oscillator using imbalanced coupling resonator for use in the microwave oven," *Appl. Sci.*, vol. 9, no. 14, p. 2887, Jul. 2019.

[3] S. Lee, S. Jeon, and J. Jeong, "Harmonic-tuned high efficiency RF oscillator using GaN HEMTs," *IEEE Microw. Wireless Compon. Lett.*, vol. 22, no. 6, pp. 318–320, Jun. 2012.

[4] J. Jeong and D. Jang, "Design technique for harmonic-tuned RF power oscillators for high-efficiency operation," *IEEE Trans. Ind. Electron.*, vol. 62, no. 1, pp. 221–228, Jan. 2015.

[5] S. Jeon, A. Suarez, and D. B. Rutledge, "Nonlinear design technique for high-power switching-mode oscillators," *IEEE Trans. Microw. Theory Techn.*, vol. 54, no. 10, pp. 3630–3640, Oct. 2006.

[6] S. Wang, Y. Zhao, X. Chen, and C. Liu, "A novel stability improvement method of S-band magnetron systems based on its anode current feature," *IEEE Trans. Microw. Theory Techn.*, vol. 72, no. 9, pp. 5530–5539, Sep. 2024.

[7] X. Chen, B. Yang, N. Shinohara, and C. Liu, "Low-noise dual-way magnetron power-combining system using an asymmetric H-plane tee and closed-loop phase compensation," *IEEE Trans. Microw. Theory Techn.*, vol. 69, no. 4, pp. 2267–2278, Apr. 2021.

[8] S. Wang, Y. Shen, C. Liao, J. Jing, and C. Liu, "A novel injection-locked S-band oven magnetron system without waveguide isolators," *IEEE Trans. Electron Devices*, vol. 70, no. 4, pp. 1886–1893, Apr. 2023.

[9] C. Liu, H. Huang, Z. Liu, F. Huo, and K. Huang, "Experimental study on microwave power combining based on injection-locked 15-kW $S$-band continuous-wave magnetrons," *IEEE Trans. Plasma Sci.*, vol. 44, no. 8, pp. 1291–1297, Aug. 2016.

[10] Z. Liu, X. Chen, M. Yang, P. Wu, K. Huang, and C. Liu, "Experimental studies on a four-way microwave power combining system based on hybrid injection-locked 20-kW S-band magnetrons," *IEEE Trans. Plasma Sci.*, vol. 47, no. 1, pp. 243–250, Jan. 2019.

[11] S. W. Shin, G. W. Choi, H. J. Kim, S. H. Lee, S. H. Kim, and J. J. Choi, "Frequency-tunable high-efficiency power oscillator using GaN HEMT," in *IEEE MTT-S Int. Microw. Symp. Dig.*, Anaheim, CA, USA, May 2010, pp. 1000–1003.

[12] S. W. Shin and J. J. Choi, "Frequency-tunable 150 W harmonic-tuned power oscillator," *Microw. Opt. Technol. Lett.*, vol. 53, no. 6, pp. 1459–1462, 2011.

[13] H. Ikeda and Y. Itoh, "2.4 GHz-band 20 W GaN-HFET VCO with frequency-tunable high power resonator," *IEEJ(C)*, vol. 140, no. 3, pp. 348–353, Mar. 2020.

[14] C. Liu, F. Tan, H. Zhang, and Q. He, "A novel single-diode microwave rectifier with a series band-stop structure," *IEEE Trans. Microw. Theory Techn.*, vol. 65, no. 2, pp. 600–606, Feb. 2017.

[15] Z. He, L. Yan, and C. Liu, "An adaptive power division strategy for nonlinear components in rectification," *IEEE Trans. Power Electron.*, vol. 39, no. 12, pp. 15436–15440, Dec. 2024.

[16] Y. Fei, Y. Hongxi, L. Ruizhu, H. Xinyang, Z. Anxue, and J. Zhonghe, "Class f and inverse class f dual modes dual bands power amplifier," in *IEEE MTT-S Int. Microw. Symp. Dig.*, Shanghai, China, Sep. 2020, pp. 1–3.

[17] A. Sheikhi and H. Hemesi, "Analysis and design of the novel class-F/E power amplifier with series output filter," *IEEE Trans. Circuits Syst. II, Exp. Briefs*, vol. 69, no. 3, pp. 779–783, Mar. 2022.

[18] K. Chen and D. Peroulis, "A 3.1-GHz class-F power amplifier with 82% power-added-efficiency," *IEEE Microw. Wireless Compon. Lett.*, vol. 23, no. 8, pp. 436–438, Aug. 2013.

[19] P. J. Tasker and J. Benedikt, "Waveform inspired models and the harmonic balance emulator," *IEEE Microw. Mag.*, vol. 12, no. 2, pp. 38–54, Apr. 2011.

[20] S.-K. Zhu, Q.-F. Cheng, H.-P. Fu, H.-F. Wu, and J.-G. Ma, "A highly efficient concurrent dual-band class-F power amplifier for applications at 1.7 and 2.14 GHz," in *IEEE MTT-S Int. Microw. Symp. Dig.*, Phoenix, AZ, USA, May 2015, pp. 1–4.

[21] B. Kim and J. Oh, "Dual-band continuous class-$F^{-1}$ power amplifier with second-harmonic suppression for harmonic radar systems," *IEEE Access*, vol. 12, pp. 62358–62364, 2024.

[22] Y.-W. Duan, J. Xu, J.-H. Su, M. Zhao, F. Liu, and G.-Q. Zhou, "Low-voltage continuous class-F wideband power amplifier," *IEEE Microw. Wireless Technol. Lett.*, vol. 34, no. 6, pp. 639–642, Jun. 2024.

[23] A. Ghanaatian, A. Abrishamifar, A. Rahmati, and A. Alizadeh, "A 15-W X-band inverse class-F GaN power amplifier with second-harmonic input tuning," *IEEE Microw. Wireless Technol. Lett.*, vol. 33, no. 9, pp. 1317–1320, Sep. 2023.

[24] Q. F. Cheng, H. P. Fu, S. K. Zhu, H. F. Wu, and J. G. Ma, "High-efficiency GaN class-F/class-$F^{-1}$ power amplifiers with distributed L-shaped parasitic-compensation circuit," *Microw. Opt. Technol. Lett.*, vol. 57, no. 10, pp. 2441–2445, 2015.

[25] M. Makimoto and S. Yamashita, "Bandpass filters using parallel coupled stripline stepped impedance resonators," *IEEE Trans. Microw. Theory Techn.*, vol. 28, no. 12, pp. 1413–1417, Dec. 1980.

[26] E. Lindberg, "The Barkhausen criterion (observation?)," in *Proc. 18th IEEE Workshop Nonlinear Dyn. Electron. Syst.*, Jun. 2010, pp. 15–18.

[27] C. Bansleben and W. Heinrich, "Electronic frequency tuning of a high-power 2.45 GHz GaN oscillator," in *IEEE MTT-S Int. Microw. Symp. Dig.*, May 2015, pp. 1–4.